\documentclass[
  ,draft            
  ]
  {aipproc}

\layoutstyle{6x9}

\begin{document}

\title{The Resummed Photon Spectrum in Radiative Upsilon Decays (And More)}

\author{Sean Fleming}{
  address={Physics Department, Carnegie Mellon University, Pittsburgh PA 15213}
}

\begin{abstract}
In this talk I present the results of two calculations that make use
of Non-Relativistic QCD and the newly developed Soft-Collinear
Effective Theory. The first process considered is inclusive
radiative $\Upsilon$ decay. The second process considered is the
leading color-octet contribution to $e^+ e^- \rightarrow J/\psi + X$. 

\end{abstract}

\maketitle


Bound states of heavy quarks and antiquarks have been of great
interest since the discovery of the $J/\psi$~\cite{Aubert:1974js,Augustin:1974xw}. 
In particular the decay and production of quarkonium is an interesting
probe of both perturbative and nonperturbative aspects of QCD
dynamics. 
A systematic theoretical framework for handling the different scales
characterizing both the decay and production of quarkonium is
Non-Relativistic Quantum Chromodynamics (NRQCD)~\cite{Bodwin:1995jh,
Luke:2000kz}. NRQCD solves important conceptual as well as
phenomenological problems in quarkonium theory. For instance,
perturbative calculations of the inclusive decay rates for $\chi_c$
mesons in the color-singlet model suffer from nonfactorizable infrared
divergences~\cite{Barbieri:1976fp, Barbieri:1981gj, Barbieri:1981xz}. 
NRQCD provides a generalized
factorization theorem so that infrared safe calculations of inclusive 
decay rates
are possible~\cite{Bodwin:1992ye}. In addition, color-octet production
mechanisms are critical for understanding the production of $J/\psi$
at large transverse momentum, $p_\perp$, at the Fermilab Tevatron
\cite{Braaten:1995vv, Cho:1996vh, Cho:1996ce}.  
There are
still many challenging problems in quarkonium physics that remain to
be solved~\cite{Bodwin:2002mr}.  One important problem is the
polarization of $J/\psi$ at the Tevatron. NRQCD predicts the $J/\psi$
should become transversely polarized as the $p_\perp$ of the $J/\psi$
becomes much larger than $2 m_c$~\cite{Cho:1995ih,Leibovich:1997pa,Beneke:1997yw,Braaten:1999qk}. 
The theoretical
prediction is consistent with the experimental data at intermediate
$p_\perp$, but at the largest measured values of $p_\perp$ the
discrepancies are at the 3$\sigma$ level~\cite{Affolder:2000nn}.
In this talk I present two additional puzzles where progress has been 
made lately: radiative $\Upsilon$ decay, and $e^+ e^- \rightarrow J/\psi + X$. 

Inclusive decays of quarkonium are understood in the framework of the
operator product expansion (OPE), with power-counting rules given by 
NRQCD. The OPE for the direct photon spectrum of $\Upsilon$ decay 
is~\cite{Bodwin:1995jh}
\begin{equation} \label{nrqcdope}
\frac{d \Gamma}{d z} = \sum_n C_n (M,z) 
   \langle \Upsilon \vert {\cal O}_n \vert \Upsilon \rangle \,,
\end{equation}
where $z = 2 E_\gamma /M$, with $M = 2 m_b$. The $C_i$ are
short-distance coefficients, and the ${\cal O}$ are NRQCD
operators . At leading order in $v$
only one term in the sum must be kept, the so called color-singlet
contribution. 

This simple picture of the photon spectrum in inclusive $\Upsilon$
decays is only valid in the intermediate range of the photon energy
spectrum ($0.3  \stackrel{<}{\sim} z  \stackrel{<}{\sim} 0.7$). In the lower range,
$z \stackrel{<}{\sim} 0.3$, photon-fragmentation contributions are
important~\cite{Catani:1995iz, Maltoni:1998nh}. At large values of the
photon energy, $ z \stackrel{>}{\sim} 0.7$, both the perturbative
expansion~\cite{Maltoni:1998nh} and the OPE~\cite{Rothstein:1997ac}
break down.

\begin{figure}\label{upsilon}
  \includegraphics[height=.35\textheight]{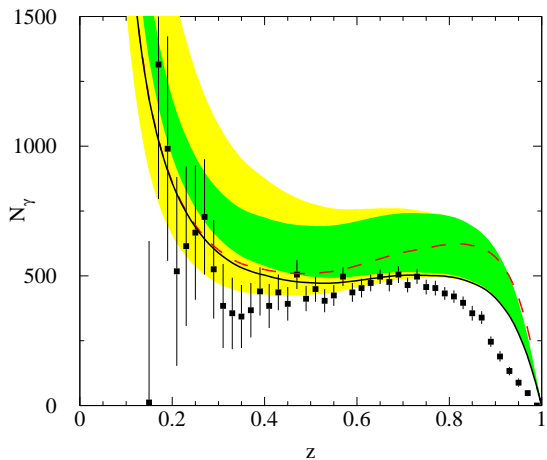}
  \caption{The inclusive radiative $\Upsilon$ photon spectrum,
compared with data from CLEO \cite{Nemati:1996xy}.}
\end{figure}

The breakdown at the endpoint is a
consequence of NRQCD not containing the correct low energy degrees of
freedom. The effective
theory which correctly describes this kinematic regime is a
combination of NRQCD for the heavy degrees of freedom, and the
soft-collinear effective theory
(SCET)~\cite{Bauer:2000ew,Bauer:2000yr,Bauer:2001ct,Bauer:2001yt} for
the light degrees of freedom. In 
Refs.~\cite{Bauer:2001rh,Fleming:2002rv,Fleming:2002sr} SCET was applied to 
radiative $\Upsilon$ decay.  A comparison of the calculation to 
CLEO~\cite{Nemati:1996xy} data is shown in Fig.~\ref{upsilon}. 

The error bars on the data are statistical only.  The dashed line is the
direct tree-level and fragmentation result,  and the solid curve is the sum of 
the interpolated resummed result and the fragmentation result.
For these two curves we used the value of $\alpha_s$ extracted
by CLEO from these data, $\alpha_s(M_\Upsilon) = 0.163$, which corresponds to
$\alpha_s(M_Z) = 0.110$ \cite{Nemati:1996xy}.  
We also show in this plot the interpolated 
resummed and fragmentation result, using
the PDG value of $\alpha_s(M_Z)$, including theoretical uncertainties,
denoted by the shaded region.  The lighter band also includes the variation, within the
errors, of the parameters for the quark to photon fragmentation
function extracted by ALEPH~\cite{Buskulic:1995au}.  

New problems have arisen as a result of recent measurements of the
spectra of $J/\psi$ produced at the $\Upsilon(4S)$ resonance in
$e^+e^-$ collisions by the BaBar and Belle
experiments~\cite{Abe:2001za, Aubert:2001pd}.  Leading order NRQCD
calculations predict that for most of the range of allowed energies
prompt $J/\psi$ production should be dominated by color-singlet
production mechanisms, while color-octet contributions dominate when
the $J/\psi$ energy is within a few hundred MeV of
the maximum allowed. Furthermore, as pointed out in
Ref.~\cite{Braaten:1996ez}, color-octet processes predict a
dramatically different angular distribution for the $J/\psi$. 

Experimental results do not agree with these expectations: the data does not 
exhibit any enhancement in the bins closest to the endpoint. However, the
total cross section measured by the two experiments exceeds
predictions based on the color-singlet model alone. The total prompt
$J/\psi$ cross section, which includes feeddown from $\psi^\prime$ and
$\chi_c$ states but not from $B$ decays, is measured to be
$\sigma_{tot} = 2.52 \pm 0.21 \pm 0.21$ pb by BaBar, while Belle
measures $\sigma_{tot} = 1.47 \pm 0.10 \pm 0.13$ pb. Estimates of the
color-singlet contribution range from $0.4 -0.9$ pb
\cite{Cho:1996cg,Yuan:1997ep,Baek:1998yf,Schuler:1998az}. Furthermore,
the angular distribution disagrees with color-singlet result. 
These aspects of the data suggest that there is a substantial
color-octet contribution which is not confined to the very endpoint. 

In Ref.~\cite{Fleming:2003gt} the endpoint region is treated within the
framework of NRQCD and SCET. The calculation depends on
a nonperturbative function, and thus is not predictive. However, moments 
of the shape function are NRQCD operators whose size is constrained by the 
velocity scaling rules of NRQCD. Choosing a simple ansatz for the shape
function whose moments are consistent with velocity scaling rules, one 
finds that the combined perturbative and nonperturbative effects lead
to substantial broadening of the color-octet spectrum in a manner that
is consistent with data. 

\begin{figure}\label{epem}
  \includegraphics[height=.3\textheight]{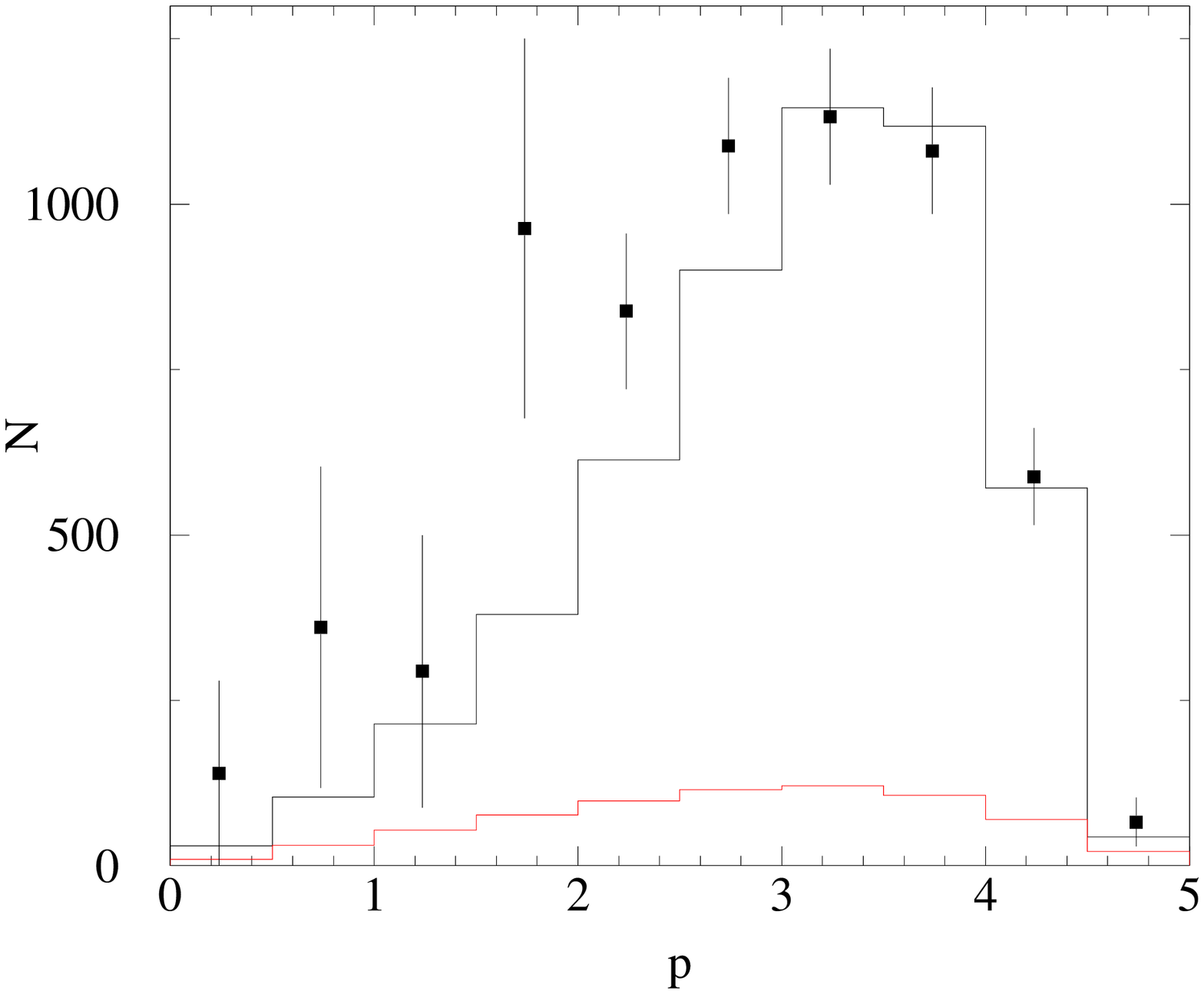}
  \includegraphics[height=.3\textheight]{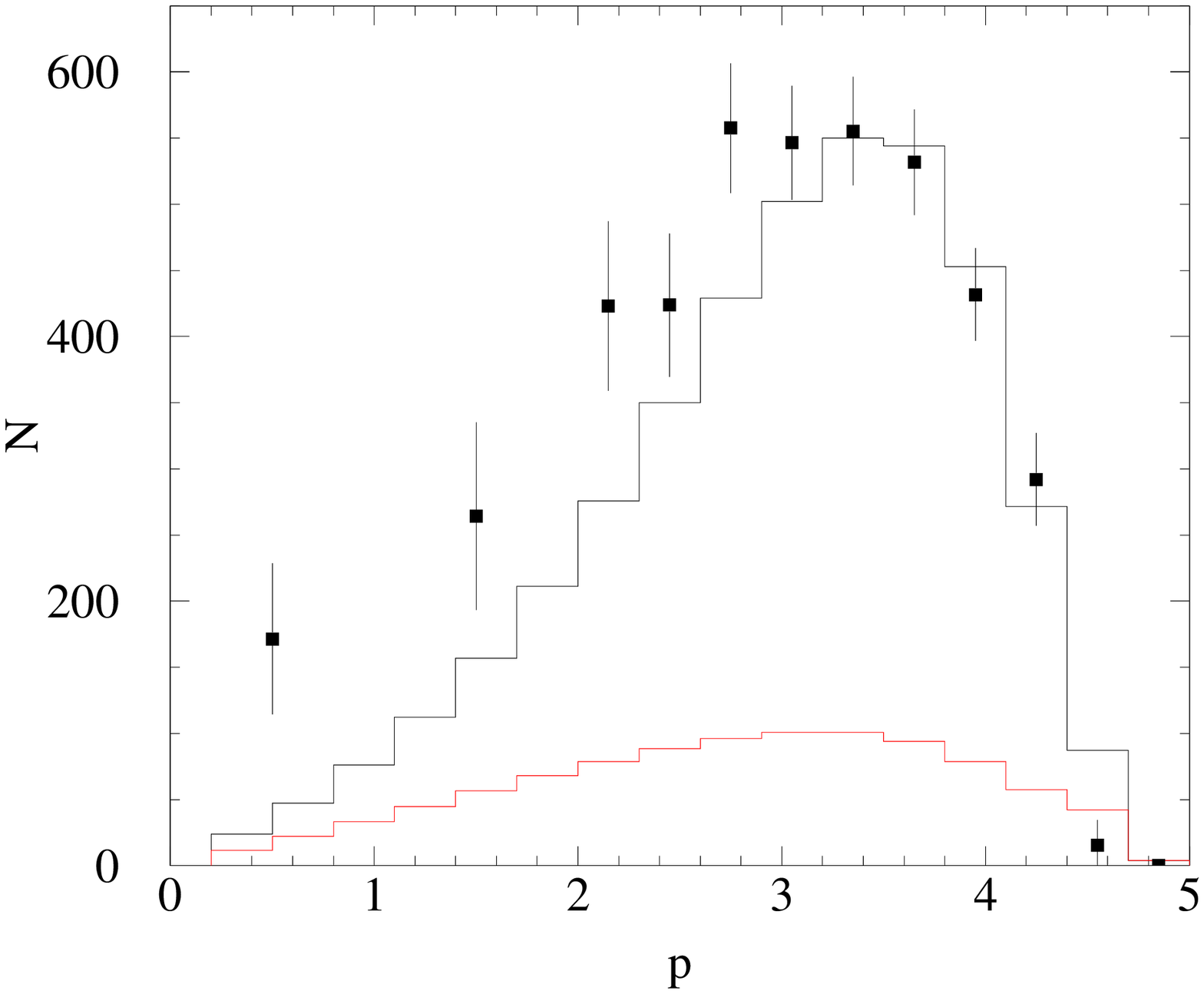}
  \caption{The sum of the color-octet and color-singlet contributions are plotted as the upper line. The lower line is the color-singlet contribution only. The graph on the left shows data from the 
BaBar collaboration~\cite{Aubert:2001pd}. The graph on the right shows data are from the Belle collaboration~\cite{Abe:2001za}. }
\end{figure}

In Fig.~\ref{epem} I show the sum of the color-octet and color-singlet contributions as the upper line, and the color-singlet contribution only as the lower line.  The color-octet matrix elements 
set the normalization. In the graph on the left they are chosen to be $\langle {\cal O}^\psi_8 (^1S_0)\rangle = \langle {\cal O}^\psi_8 (^3P_0)\rangle/m_c^2 = 1.3 \times 10^{-1}{\rm\
GeV}^3$. This is plotted against the BaBar data~\cite{Aubert:2001pd}. In the graph on the right they are chosen to be $\langle {\cal O}^\psi_8 (^1S_0)\rangle = \langle {\cal O}^\psi_8 (^3P_0)\rangle/m_c^2 = 6.6 \times 10^{-2}{\rm\ GeV}^3$, and is plotted against the Belle data~\cite{Abe:2001za}. 

While the calculations of Ref.~\cite{Fleming:2003gt} show that the leading color-octet
contribution is broad enough to be compatible with the observed
$p_\psi$ distributions, other features of the $e^+e^-$ data remain
puzzling. In particular, Belle reports a large ratio of $J/\psi + c \bar{c}$ over inclusive 
$J/\psi$~\cite{Abe:2002rb}.
The predicted ratio from leading order color-singlet production
mechanisms alone is at least a factor of three too small \cite{Cho:1996cg,Baek:1998yf} 
and a large color-octet contribution makes this ratio even smaller.

\begin{theacknowledgments}
 I would like to thank my collaborators Adam Leibovich and Tom Mehen. 
 This work is  supported  by Department of Energy grant number DOE-ER-40682-143.
\end{theacknowledgments}


\bibliographystyle{aipproc}   

\bibliography{fleming}

\IfFileExists{\jobname.bbl}{}
 {\typeout{}
  \typeout{******************************************}
  \typeout{** Please run "bibtex \jobname" to optain}
  \typeout{** the bibliography and then re-run LaTeX}
  \typeout{** twice to fix the references!}
  \typeout{******************************************}
  \typeout{}
 }

\end{document}